\documentclass[cits]{PoS}

\usepackage{url}

\title{Quantification of nuclear uncertainties in nucleosynthesis of elements beyond Iron}

\ShortTitle{Quantification of nuclear uncertainties in nucleosynthesis of elements beyond Iron}

\author{\speaker{Thomas Rauscher}\\
        Centre for Astrophysics Research, University of Hertfordshire\\
Hatfield AL10 9AB, United Kingdom\\
and\\
Department of Physics, University of Basel\\
4056 Basel, Switzerland\\
and\\
UK Network for Bridging Disciplines of Galactic Chemical Evolution (BRIDGCE)\\
\url{http://www.astro.keele.ac.uk/bridgce}, United Kingdom
}

\abstract{Nucleosynthesis beyond Fe poses additional challenges not encountered when studying astrophysical processes involving light nuclei. Generally higher temperatures and nuclear level
densities lead to stronger contributions of transitions on excited target states. This may prevent cross section measurements to determine stellar reaction rates and theory contributions remain important. Furthermore, measurements often are not feasible in the astrophysically relevant energy range. Sensitivity analysis allows not only to determine the contributing nuclear properties but also is a handy tool for experimentalists to interpret the impact of their data on predicted cross sections and rates. It can also speed up future input variation studies of nucleosynthesis by simplifying an intermediate step in the full calculation sequence. Large-scale predictions of sensitivities and ground-state contributions to the stellar rates are presented, allowing an estimate of how well rates can be directly constrained by experiment. The reactions $^{185}$W(n,$\gamma$) and $^{186}$W($\gamma$,n) are discussed as application examples. Studies of uncertainties in abundances predicted in nucleosynthesis simulations rely on the knowledge of reaction rate errors. An improved treatment of uncertainty analysis is presented as well as a recipe for combining experimental data and theory to arrive at a new reaction rate and its uncertainty. As an example, it is applied to neutron capture rates for the s-process, leading to larger uncertainties than previously assumed.}

\FullConference{XIII Nuclei in the Cosmos\\
                 7-11 July, 2014\\
                 Debrecen, Hungary}

\begin{document}

\section{Introduction}

\begin{figure}
\includegraphics[width=\columnwidth]{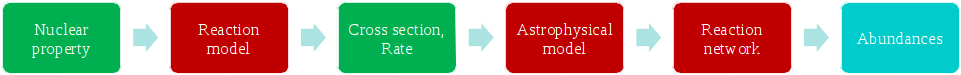}
\caption{\label{fig:propagation}Calculation sequence for nucleosynthesis studies, implying the sequence of error propagation.}
\end{figure}

\noindent
Considering the fact that nuclear physics input to astrophysical simulations still bears considerable uncertainty, both because of the involvement of unstable nuclides and tiny reaction cross sections unprobeable in the laboratory, the question of what range of astrophysical results is allowed within the given uncertainties remains important. Of interest is the impact of uncertainties in nuclear properties required for cross section calculations, reaction cross sections, and finally astrophysical reaction rates. All of these feed into each other (see Fig.\ \ref{fig:propagation}) and any error at any stage will affect the subsequent calculations. Errors in each step not only arise through the uncertainties in the input but also in the theoretical model used to predict the desired quantity, which is inplementing input of other calculations or experimental data.
Closely connected is the question of how strongly laboratory measurements can actually constrain an astrophysical rate (\textit{stellar} rate). Only recently, first attempts have been made to determine these limits \cite{rau12}. As will be argued below, an experimental error cannot be directly adopted as a rate uncertainty but further considerations have to be included, often leading to larger uncertainties in the rates than the experimental ones. 
Finally, the impact of rate uncertainties on predicted abundances has to be explored. This can be done by identification of major reaction flows in a reaction network or by various types of Monte Carlo variation studies. The study of reaction flows is only helpful when few major reactions contribute and an unique reaction path can be found. Especially nucleosynthesis around and above the Fe-group involves large reaction networks for which often this is not possible, Then a Monte Carlo analysis is the only way to quantify uncertainties in the final abundances (see, e.g., \cite{raunicposter,nobnicposter}).

\section{Uncertainties in Nucleosynthesis Calculations and Model Sensitivities}
\label{sec:sensi}

\noindent
In each step shown in Fig.\ \ref{fig:propagation}, two fundamentally different sources of errors have to be distinguished: Type I uncertainties are the usual statistical errors as used in measurements, type II errors are those stemming from an inappropriate choice of theoretical model. While statistical errors can be well quantified and propagated, this is not possible for type II errors because even a comparison of various models does not provide a systematic, statistically valid sample of possibilities to model the quantities in question \cite{rau12a}. While type I errors in input quantities can be propagated analytically or by a \emph{systematic} variation in the input, this is ruled out for type II errors which are not quantifiable. This becomes a limitation for a final result as well as for input based on some theoretical description. This has to be borne in mind whenever uncertainties in calculations are discussed.

It would be desireable to propagate the uncertainties through the full calculation sequence shown in Fig.\ \ref{fig:propagation}. As this is not yet feasible, one of two types of variation studies is usually performed: variation of input to nuclear reaction models to arrive at a combined uncertainty of the predicted reaction cross sections or variations of stellar reaction rates in reaction network calculations to estimate the uncertainty in the final abundances. With advances in computing power, Monte Carlo variations have become increasingly popular in performing such studies (e.g., \cite{raunicposter,nobnicposter,mohrNIC,coc}).

\emph{Model sensitivities} can help to disentangle input and model uncertainties and speed up the treatment of error propagation through the calculation chain as their utilization removes the need to run a full model calculation for each set of varied input. In combination with fast, parallelized Monte Carlo frameworks, this may allow a move to more comprehensive uncertainty studies in the future. Sensitivities also allow experimentalists to easily assess impacts of measured quantities (either in connection with already performed measurements or for planning of future experiments) without the need to perform a full model calculation.
The sensitivity $s$ of a derived quantity $\Omega$ to a change in an input quantity $q$ is defined as \cite{rau12a,rau14} $s=(v_\Omega-1)/(v_q-1)$.
It is a measure of a change by a factor of $v_\Omega=\Omega_\mathrm{new}/\Omega_\mathrm{old}$ in $\Omega$ as the result of a change in the quantity $q$ by the factor $v_q=q_\mathrm{new}/q_\mathrm{old}$, with $s=0$ when no change occurs and $s=1$ when the final result changes by the same factor as used in the variation of $q$, i.e., $s=1$ implies $v_\Omega=v_q$.
This is equivalent to writing
$s=(q_\mathrm{old}/\Omega_\mathrm{old})(d\Omega/dq)$,
with $d\Omega=\Omega_\mathrm{new}-\Omega_\mathrm{old}$ and $dq=q_\mathrm{new}-q_\mathrm{old}$, as used in standard sensitivity analysis. Derived quantities $\Omega$ of interest here are reaction cross sections and rates but in principle could be also final abundances from a nucleosynthesis model.

\begin{figure}
\includegraphics[width=0.7\columnwidth]{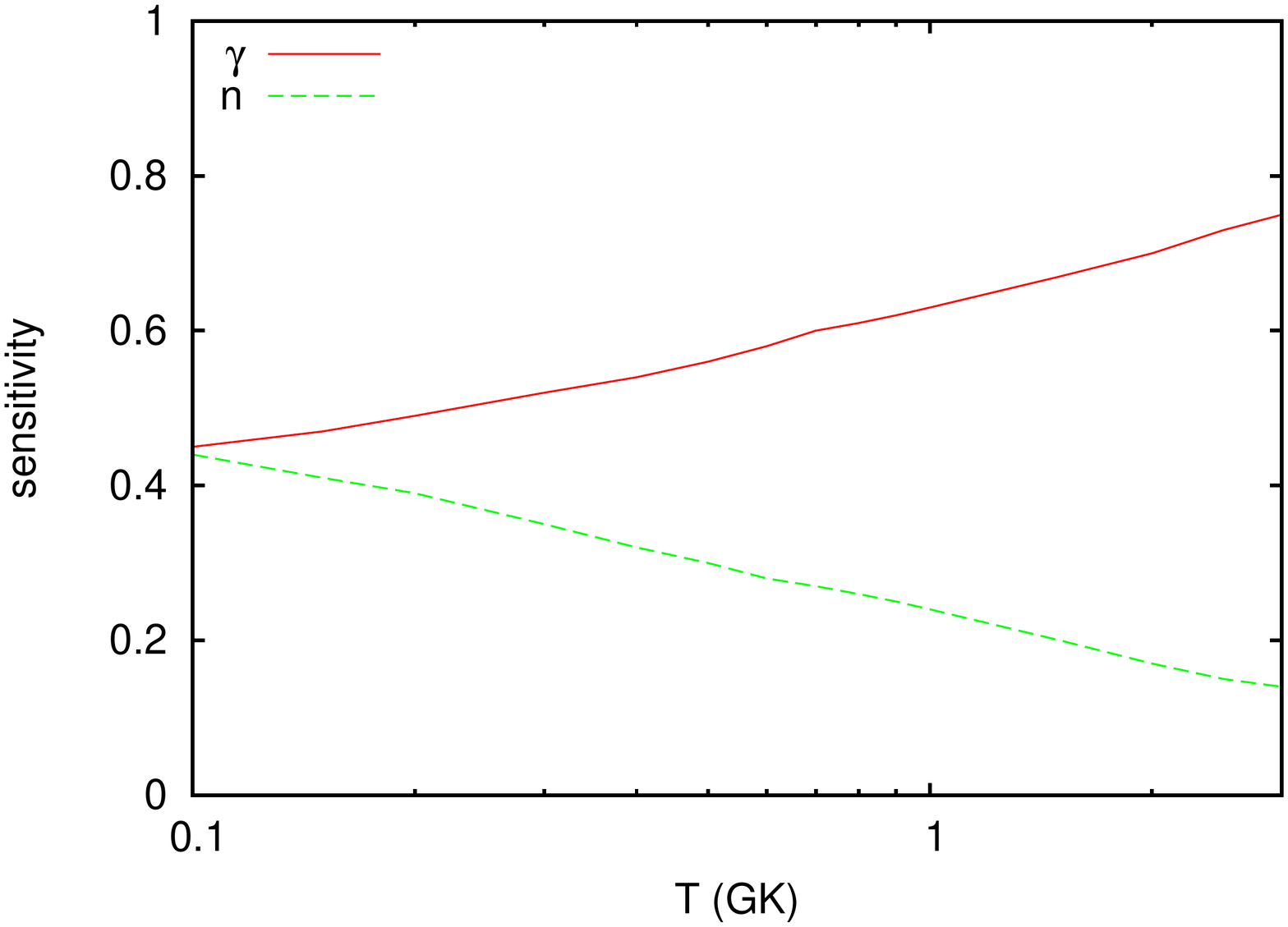}
\caption{Absolute values of the sensitivity $|s|$ of the stellar $^{185}$W(n,$\gamma$) rate to variations in the neutron and $\gamma$ widths as function of plasma temperature \cite{rau12a,rau14}.\label{fig:w185}}
\end{figure}

Using sensitivities is useful when a model depends on a limited number of input quantities. This would be the case, for example, for compound nucleus reactions which depend on transmission coefficients and the reaction widths derived from them \cite{raurev}. Tables of rate and cross section sensitivities for compound reactions between Ne and Bi from p-drip to n-drip have been provided in \cite{rau12a}. To directly infer the impact of a (experimentally or theoretically) newly determined quantity $q$ (for example, an averaged width in a Hauser-Feshbach calculation) on the cross section or reaction rate, the relation $\Omega_\mathrm{new}=\Omega_\mathrm{old} (s\times (\overline{v}_q-1)+1)$ can be used \cite{rau14},
with $\Omega_\mathrm{old}$ being the previous value of the cross section or rate of interest,  $\Omega_\mathrm{new}$ being the new value, and $\overline{v}_q$ the factor by which the newly determined $q$ differs from its previous value used to calculate $\Omega_\mathrm{old}$. This may be especially useful for experimentalists and also disentangles comparisons to data from the theory discussion of changes in reaction widths which may also depend on nuclear models.

\section{Stellar rates and contributions from excited states}
\label{sec:stellrates}

\noindent
Rates and cross sections exhibit a different sensitivity to input variations because of the additional transitions between excited states appearing in \emph{stellar} reaction rates \cite{raurev,rau12lett}.
The stellar reaction rate $r^*$ at plasma temperature $T$ is given by \cite{rau14,raurev}
\begin{equation}
\frac{r^*(T)}{n_a n_A} = \sqrt{\frac{8}{\pi m_{aA}}} \left(kT\right)^{-3/2} \int_0^\infty \sigma^*(E,T) E e^{-E/(kT)}\,dE \quad,
\label{eq:stellrate}
\end{equation}
where $n_a$, $n_A$ are the number densities of projectiles and target nuclei, respectively, $m_{aA}$ is the reduced mass, and $\sigma^*(E,T)$ is an energy- and temperature-dependent \emph{stellar} reaction cross section, 
\begin{equation}
\sigma^*(E,T)=\frac{1}{G(T)} \sum_i \sum_j (2J_i+1) \frac{E-E_i}{E}
\sigma^{i \rightarrow j}(E-E_i) = \frac{1}{G(T)} \sum_i \sum_j W_i \sigma^{i \rightarrow j}(E-E_i)
\quad.
\label{eq:stellcs}
\end{equation}
This means that the partial cross sections $\sigma^{i \rightarrow j}$ for reactions on excited states are evaluated at $E-E_i$ instead of the usual c.m.\ energy $E$ and their contributions weighted by $W_i$. Cross sections for individual transitions $\sigma^{i \rightarrow j}$ are zero for negative energies \cite{fow}. The nuclear partition function is denoted by $G(T)$.
It should be noted that the relative weights $W_i$ of the excited state contributions to the stellar cross section are linearly decreasing instead of the exponential decline seen in the Boltzmann population factors $P_i=(2J_i+1)\exp(-E_i/(kT))$. This leads to a larger contribution of target states at higher excitation energy than expected from the $P$ alone. It remains true, nevertheless, that contributions of excited target state transitions are more important in heavier nuclei, exhibiting a higher nuclear level density, than in light systems \cite{raurev}.
The contribution of state $i$ with spin $J_i$ and excitation energy $E_i$ to the stellar rate $r^*$ can be quantified as \cite{rau12a,rau12lett}
\begin{equation}
\label{eq:xfactor}
X_i(T)=\frac{2J_i+1}{G(T)}e^{-E_i/(kT)}\frac{\int\sigma_i(E)E e^{-E/(kT)}\,dE}{\int\sigma^*(E,T)E e^{-E/(kT)}\,dE} \quad,
\end{equation}
where the reaction cross section $\sigma_i$ is given by
$\sigma_i(E)=\sum_{j}\sigma^{i\rightarrow j}(E-E_i)$.
It is very important to note that the ground state (g.s.) contribution $X_0$ is different to the simple ratio $r_0/r^*$ of g.s.\ and stellar rates, respectively \cite{rau12}. The ratio $r^*/r_0$ is called the \emph{stellar enhancement factor} (SEF) and in the past was mistakenly assumed to quantify the excited state contributions.
Using the above definition, the total excited state contribution becomes $X_\mathrm{exc}=1-X_0$. The $X_0$ are tabulated in \cite{rau12a,rau12lett}.

Current laboratory measurements determine reaction cross sections of target nuclei in their g.s.
Combining the information contained in sensitivities and g.s.\ contributions allows to assess how useful a measurement is to derive a stellar rate. For example, the nucleus $^{185}$W is unstable and the location of a branching in the $s$-process path where neutron capture and $\beta$ decay are competing. 
Measuring $^{185}$W(n,$\gamma$) would provide a strong experimental constraint on the stellar rate because $X_0=0.98$ at $kT=8$ keV and $X_0=0.75$ at $kT=30$ keV, at two important s-process energies. Due to the instability of $^{185}$W, however, such a measurement has not been feasible, yet. Instead, $^{186}$W($\gamma$,n) has been measured in a photodisintegration experiment (see references in \cite{rau14}). For this reaction, however, $X_0=7\times 10^{-3}$ at $T_9=0.1$ and $X_0=5\times 10^{-3}$ at $T_9=0.4$, for instance. The measured contribution to the stellar rate is negligible, which is typical for photodisintegration measurements \cite{rau12a}.
The tiny $X_0$ for photodisintegration are closely related to the relevance of the measured $\gamma$ transition to/from the g.s.\ of $^{186}$W. Among the $\gamma$s emitted in the capture, the ones directly to the g.s.\ are only a tiny fraction of all \cite{rau14,raurev}. The ($\gamma$,n) experiment probes the $\gamma$-strength function at an energy which is important neither for the capture nor for the stellar ($\gamma$,n) rate. Even if the $\gamma$ width was constrained, however, this would still not fully constrain the (n,$\gamma$) rate. As shown in Fig.\ \ref{fig:w185}, the rate is sensitive to uncertainties in the $\gamma$- as well as the neutron width at low temperatures. It follows that the $^{185}$W(n,$\gamma$) rate still is experimentally unconstrained at $s$-process temperatures.

\section{How to combine theory and measurement in a revised stellar rate}

\noindent
The introduction of g.s.\ contributions allows an improved treatment of the question how experimental and theoretical information can be combined to arrive at a better constrained stellar rate and its new uncertainty. As discussed in \cite{rau12lett} one of two approaches has to be adopted: 1) include only what has been measured without further assumptions or
2) include theoretical considerations concerning correlations between $\sigma_0$ and $\sigma_{i>0}$.

\begin{figure}
\includegraphics[width=0.75\columnwidth]{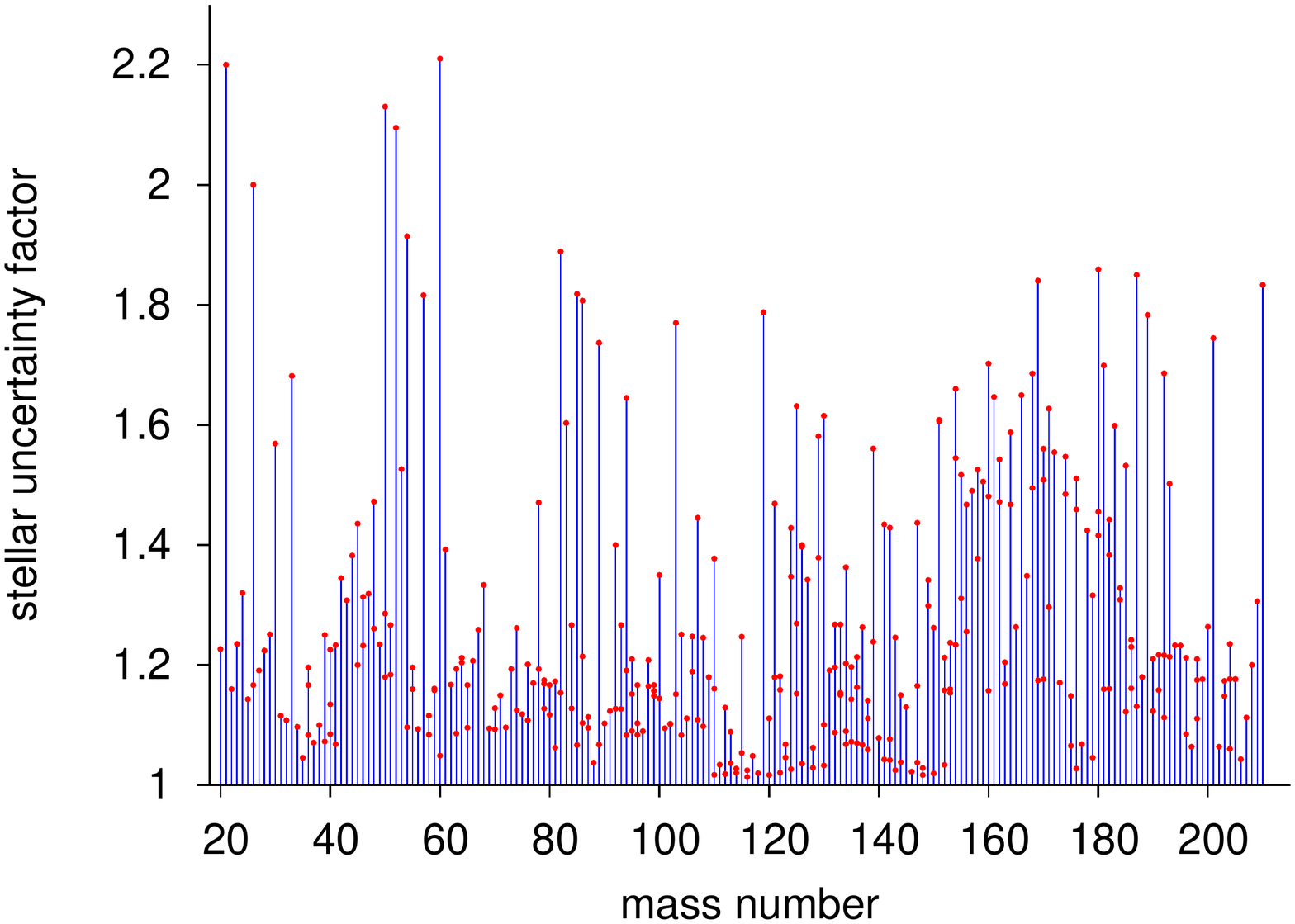}
\caption{Uncertainty factors $U^*_\mathrm{new}$ for stellar (n,$\gamma$) rates at $kT=30$ keV \cite{rau12lett}. For many cases these are considerably larger than the experimental uncertainties which are of the order of 5\% or lower.\label{fig:uncert}}
\end{figure}

Using approach 1 with the experimentally determined g.s.\ rate $r_0^\mathrm{exp}$ (by integration of $\sigma_0^\mathrm{exp}$) a correction factor $f^*(T)$ has to be applied to the previous (purely theoretical) stellar rate $r^*_\mathrm{th}$ to obtain the new rate $r^*_\mathrm{new}(T)=f^*(T) r^*_\mathrm{th}(T)$. The factor $f^*(T)=1+X_0(T)(r_0^\mathrm{exp}(T)/r_0^\mathrm{th}(T)-1)$ contains the ratio between experimental and theoretical g.s.\ rate \cite{rau12lett}.
The new uncertainty factor $u^*$ of the stellar rate is constructed from the original theory uncertainty factor $U_\mathrm{th}$ and the experimental uncertainty factor $U_\mathrm{exp}$ by using
$u^*=U_{\mathrm{exp}}+(U_\mathrm{th}-U_{\mathrm{exp}})X_\mathrm{exc}$.
This approach has been applied to neutron capture rates for the s-process which are generally assumed to be strongly constrained by high-precision measurements \cite{rau12lett}. As shown in Fig.\ \ref{fig:uncert}, the resulting uncertainties are considerably larger than previously assumed. This has important consequences for the interpretation of astrophysical results. For example, it was shown that it was only possible to explain observed Eu isotope ratios in stars and presolar grains when allowing for rates and rate uncertainties as obtained with this approach but not when only considering experimental uncertainties \cite{avila,euproceedings}.

Approach 2 includes additional theory assumptions on the factor by which
excited state contributions are renormalized in addition to replacing the g.s.\ contribution by a measured value. An extreme additional assumption would be to use the same renormalization as for the g.s.\ rate. This implies two strong restrictions: (a) the cause of any discrepancy between $r_0^\mathrm{th}$ and $r_0^\mathrm{exp}$ also causes a similar deviation of the same magnitude in all $r_{i>0}^\mathrm{th}$, and (b) there are no further uncertainties in predicted excited state transitions. If both apply, then $f^*=r_0^\mathrm{exp}/r_0^\mathrm{th}$. It has to be noted that this is equivalent to multiplying the measured g.s.\ rate $r_0^\mathrm{exp}$ by the SEF.
The uncertainty for the new stellar rate obtained in approach 2 is difficult to quantify. Only if both restrictions (a) and (b) apply, and only then, it is just the uncertainty of the measurement. If this is not the case, a different renormalization has to be applied to each excited state transition and the uncertainty will be larger.

The above two approaches are the two extreme cases. In the absence of a further detailed theoretical investigation of the uncertainty sources, it is preferrable to apply the ``pessimistic view'' of approach 1.
Errors in the predictions of $\sigma_{i>0}$ will be different than those of the $\sigma_0$ for several reasons. Equation (\ref{eq:stellcs}) shows that in $\sigma^*$ partial cross sections are evaluated at relative interaction energies $E-E_i$. Sensitivities $s$ (Sec.\ \ref{sec:sensi}) of the $\sigma^{i \rightarrow j}$ are strongly energy-dependent and therefore transitions from states $i>0$ (occurring at lower relative energy) may be sensitive to different nuclear properties than those from $i=0$. For example, for neutron captures in the $s$-process this can be important for nuclei with large capture $Q$-value and high nuclear level densities, for which the neutron widths become comparable to or smaller than the $\gamma$-widths within the covered energy range. Even if sensitivities to the various widths are not changing, different spins and parities of the excited states imply different angular momentum barriers in particle transitions, and also may give rise to a different selection of electromagnetic multipolarities in $\gamma$-transitions. This may be more important in nuclei with low level densities. Finally, it was already mentioned in Sec.\ \ref{sec:stellrates} that the prediction of low-energy $\gamma$ transitions bears a different uncertainty than the one of widths at high $\gamma$ energies.
The actual circumstances and sensitivities will be different in each case and have to be thoroughly investigated for each reaction separately. For details see \cite{rau12,rau12a,rau14}.

This work is partially supported by the Swiss NSF and the European Research Council (grant GA 321263-FISH).

\end{document}